\documentclass[runningheads]{llncs}
\usepackage[T1]{fontenc}
\usepackage{graphicx}

\usepackage{placeins}

\newcommand{\figref}[1]{\figurename~\ref{#1}}

\usepackage{hyperref}

\begin{document}
\title{
A generalisable head MRI defacing pipeline:
Evaluation on 2,566 meningioma scans
}
\titlerunning{A generalisable head MRI defacing pipeline}
%
\author{Lorena Garcia-Foncillas Macias\inst{1}
\and
Aaron Kujawa\inst{1}
\and
Aya Elshalakany\inst{1,2}
\and
Jonathan Shapey\inst{1,2}
\and
Tom Vercauteren\inst{1}
}
\authorrunning{L. Garcia-Foncillas Macias et al.}
%
\institute{School of Biomedical Engineering \& Imaging Sciences, King’s College London \and Department of Neurosurgery, King’s College Hospital NHS Foundation Trust}
\maketitle              
\begin{abstract}
Reliable MRI defacing techniques to safeguard patient privacy while preserving brain anatomy are critical for research collaboration. 
Existing methods often struggle with incomplete defacing or degradation of brain tissue regions.
We present a robust, generalisable defacing pipeline for high-resolution MRI that integrates atlas-based registration with brain masking. Our method was evaluated on 2,566 heterogeneous clinical scans for meningioma and achieved 99.92\% success rate (2,564/2,566) upon visual inspection.
Excellent anatomical preservation is demonstrated with a Dice similarity coefficient of \( 0.9975 \pm 2.3 \times 10^{-3} \) between brain masks automatically extracted from the original and defaced volumes. Source code at 
\url{https://github.com/cai4cai/defacing_pipeline}.
\end{abstract}

\subsubsection{Purpose} 

\begin{figure}[t]
\centering
\includegraphics[width=\textwidth]{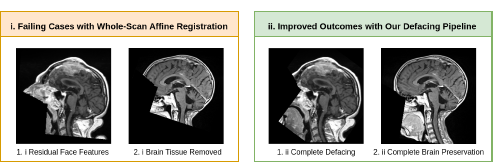}
\caption{Comparison of (i) defacing failures with affine registration and (ii) corrected cases using the proposed pipeline.}
\label{fig:cases}
\end{figure}

The availability of reliable defacing techniques to protect patient identity is increasingly important for cross-institutional research collaborations.
Existing methods include:
1)~registration-based approaches using predefined atlases such as \texttt{AFNI} \texttt{Refacer}~\cite{Cox1996}, \texttt{mri\_reface} \cite{mridefacer}, and \texttt{PyDeface}~\cite{pydeface};
2)~geometry-based techniques including \texttt{QuickShear}~\cite{Schimke2011};
and 3)~model-based frameworks such as \texttt{DeepDefacer}~\cite{Khazane2022}.
While widely adopted, registration-based methods can fail to achieve accurate alignment, leaving facial features or removing brain tissue. 
\texttt{QuickShear} provides a geometry-driven alternative by using a brain mask to compute the convex hull on a middle slice, from which a defacing plane is derived. However, this approach does not guarantee full preservation of brain tissue, limiting its reliability. Additionally, model-based techniques may lack generalisability across different sequences or pathological cases. 
Preliminary experiments illustrated in \figref{fig:cases} with currently available tools on a large-scale clinical dataset highlighted the need for an alternative with higher success rate.

To address limitations of widely available tools, this study presents the development of a robust, generalisable defacing pipeline for high-resolution MRI data. 
The proposed method combines atlas-based registration with brain masking to reliably remove facial features while preserving brain anatomy. It maintains the native image space and voxel dimensions, ensuring the defaced data remain suitable for downstream quantitative analyses.

\subsubsection{Material and methods}
\begin{figure}[t]
\includegraphics[width=\textwidth]{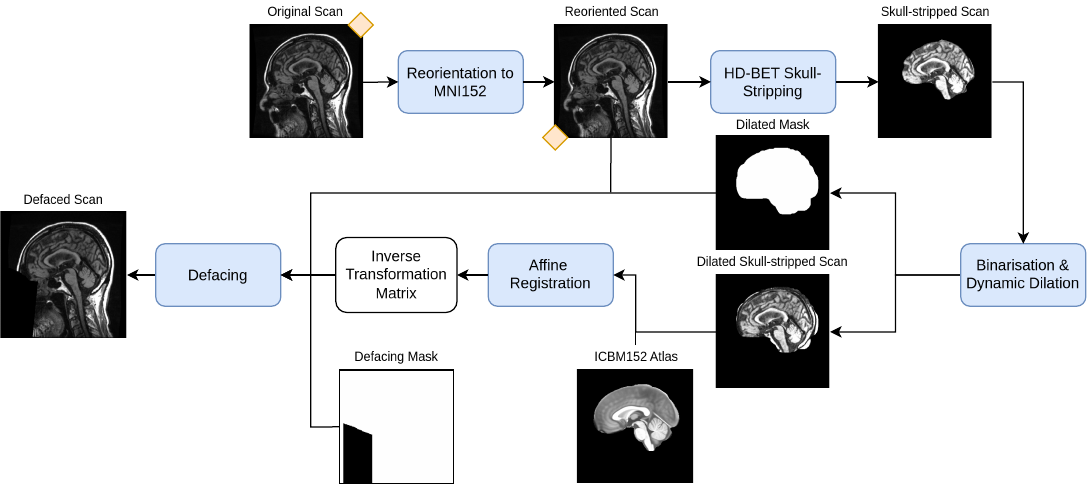}
\caption{Overview of the proposed MRI defacing pipeline. Steps are shown in blue; the rhomboid illustrates voxel orientation before and after reorientation to MNI152 space.}
\label{fig:pipeline}
\end{figure}

The proposed open-source pipeline~\cite{github} illustrated in \figref{fig:pipeline} comprises the following stages: 
(1) reorientation using \texttt{fslreorient2std} \cite{Smith2004} to match the orientation of the 
MNI152 template;
(2) tight skull-stripping using \texttt{HD-BET}~\cite{Isensee2019}; 
(3) binarisation of the tightly skull-stripped images to obtain a binary brain mask; 
(4) morphological dilation of the brain mask with a safety margin of 7 mm to preserve relevant features near the brain boundary such as potential convexity meningiomas; 
(5) generation of a loosely skull-stripped image by applying the dilated brain mask to the original image;
(6) affine registration of the loosely skull-stripped image to a skull-stripped T1-weighted 
ICBM152
reference template~\cite{icbm152_nist,fonov2011unbiased} using mutual information as the similarity metric; 
(7) application of the inverse transformation matrix to a defacing mask, followed by resampling to generate an initial registered defacing mask; 
(8) union of the initial registered defacing mask and the dilated brain mask to generate a \emph{brain-safe} defacing mask;
and (9) generation of the final defaced output.

The pipeline was evaluated on a large, heterogeneous dataset comprising 4,932 MRIs obtained from King’s College Hospital from 105 newly diagnosed meningioma patients, across 453 scanning sessions between January 2012 and November 2020. 
Ethical approval was acquired from the NHS Health Research Authority Research Ethics Committee (REC Ref 22/NS/0160). 
After excluding localisers, diffusion, perfusion, angiography, and other non-structural sequences,
2,566 scans were retained for defacing.
The dataset included a wide range of scanners, acquisition protocols, image resolutions, and anatomical coverage.

A biomedical engineer with 3 years of experience performed quality control of all scans to confirm complete removal of identifiable facial features (mouth, nose, eyes) and absence of alteration of the brain. 
As an additional automated measure to evaluate whether the defacing algorithm successfully removed facial features without compromising brain tissue, the Dice Similarity Coefficient (DSC) was calculated between the brain masks produced by \texttt{HD-BET} from 1) the original and 2) defaced scans. 
Similarly, anatomical label propagation was conducted on the original and defaced images using a single-atlas approach with the CerebrA atlas~\cite{Manera2020}. Subsequently, DSCs were computed for the resulting segmentations to evaluate the impact of defacing on the accuracy of anatomical labelling. 

\subsubsection{Results}
Quality control revealed a 99.92\% success rate. 
Out of the 2,566 scans, only two demonstrated suboptimal defacing, with residual facial features. 
Both failures were fat-only images acquired using the Turbo Spin Echo (TSE) Dixon sequence~\cite{ismrm2016_tse_dixon}.
We hypothesise that, although \texttt{HD-BET} successfully segmented the brain region in these cases, the imaging characteristics affected the affine registration stage, where mutual information was used as the similarity metric, resulting in less accurate alignment than for other sequences.

The average DSC obtained from comparing the binary brain masks produced by \texttt{HD-BET} for the original and defaced images was \( 0.9975\) with a standard deviation of \(2.3 \times 10^{-3} \). 
This near-perfect overlap confirms that the defacing process reliably preserved brain tissue and further demonstrates that models such as \texttt{HD-BET} can accurately identify and segment the brain post-defacing. 
From the single-atlas label propagation, DSCs across scans for individual labels were consistently high, with the best-performing label achieving an average of \( 0.9999 \pm 4.0 \times 10^{-5} \), and the lowest-performing one obtaining \( 0.9997 \pm 8.3 \times 10^{-4} \). 
These results indicate that the defacing pipeline preserves brain regions sufficiently for consistent brain mask generation and anatomical labelling accuracy using \texttt{HD-BET} and single-atlas propagation.

\subsubsection{Conclusion}
Integrating skull-stripping and registration using this proved fundamental to the success of our defacing pipeline as shown in \figref{fig:cases}.
This strategy ensures that registration is performed relative to the brain itself
while the accompanying brain mask reliably prevents inadvertent removal of brain tissue during defacing. The resulting pipeline enables secure, multi-centre sharing of sensitive neuroimaging data while retaining anatomical integrity for downstream applications. Future work will evaluate the preservation of individual brain structures through multi-atlas label propagation.


\FloatBarrier
\begin{credits}
\subsubsection{\ackname} LGFM is supported by the EPSRC CDT in Smart Medical Imaging [EP/S022104/1].

\end{credits}
%
%
%
\bibliographystyle{splncs04}

\end{document}